\documentclass[epjCONF,columns]{svjour}
\usepackage{graphics}
\usepackage[varg]{txfonts} 
\usepackage[latin1]{inputenc}

\newcommand{\beq}{\begin{equation}}
\newcommand{\eeq}{\end{equation}}
\newcommand{\bea}{\begin{eqnarray}}
\newcommand{\eea}{\end{eqnarray}}
\newcommand{\eps}{\varepsilon}

\session-title{International Conference Nuclear Structure and Related Topics Dubna,
Russia, July 2 - July 7, 2012}
\begin{document}
\title{Self-consistent calculations of quadrupole moments of spherical nuclei}
\author{S. Kamerdzhiev\inst{1}\fnmsep\thanks{\email{kamerdzhiev@ippe.ru}} \and S. Krewald \inst{2}
\and S. Tolokonnikov \inst{3,4} \fnmsep  \and  E.E. Saperstein
\inst{3} \and D.Voitenkov \inst{1}}
\institute {Institute for Physics and Power Engineering, 249033
Obninsk, Russia \and  Institut f\"ur Kernphysik, Forschungszentrum
J\"ulich, D-52425 J\"ulich, Germany \and Kurchatov Institute, 123182
Moscow, Russia \and Moscow Institute of Physics and Technology,
123098 Moscow, Russia}
\abstract{
The self-consistent Theory of Finite Fermi Systems based on the
Energy Density Functional by Fayans {\it et al}. with the set DF3-a
of parameters fixed previously is used to calculate  three
kinds of quadrupole moments. At first, we examined systematically
quadrupole moments of odd neighbors of semi-magic lead and tin
isotopes and $N=50,N=82$ isotones. Second, we found quadrupole
moments of the first $2^+$ states in the same two chains of
isotopes. Finally, we evaluated quadrupole moments of odd-odd nuclei
neighboring to double magic ones. Reasonable agreement with
available  experimental data has been obtained. Predictions are made
for quadrupole moments of nuclei in the vicinity of unstable magic
nuclei}
%
\maketitle
\section{Introduction}
\label{intro}

As the result of quick development of
experimental techniques in nuclear physics, the bulk 
data on nuclear static moments  has become very extensive and
comprehensive \cite{stone}, thus  creating a challenge to nuclear theory.
First of all,
it concerns the nuclei distant from the $\beta$-decay stability
valley
which are often close to the drip lines and are of great
interest to nuclear astrophysics. For this reason, a
theoretical approach used for describing such nuclei should
have a
high predictive power. The self-consistent Theory of Finite
Fermi Systems (TFFS) \cite{khodelsap} based on the EDF by Fayans
{\it et al.} \cite{Fay} is one of  such approaches.

A good description of the
quadrupole \cite{BE2,QM} and
magnetic\cite{Tol-Sap,mu2} moments of odd
semi-magic
nuclei has been achieved  within this
approach.
For quadrupole moments, we use a new DF3-a version 
\cite{Tol-Sap1} of the original DF3 functional \cite{Fay}
which was employed in calculations of magnetic moments. It
differs from the DF3 version only in  the spin-orbit parameters
$\varkappa,\varkappa'$ and the effective tensor force.
The DF3-a functional  is characterized
by a  rather strong effective tensor force.

In these calculations, the so-called ``single-quasiparticle
approximation'' has been used, where one quasiparticle in
the fixed state $\lambda=(n,l,j,m)$ with the energy
$\varepsilon_{\lambda}$ is added to the even-even core.
According to the TFFS \cite{AB},  a quasiparticle  differs from
 a particle of the single-particle model in two respects.
First, it possesses the local charge $e_q$  and, second, the core is polarized
 due to the interaction between the particle and the core nucleons via the
 Landau--Migdal (LM) amplitude. In other words, the quasiparticle possesses the
 effective charge $e_{\rm eff}$ caused by the polarizability  of the core, which is found
 by solving the TFFS equations.
 In the many-particle Shell Model,
 a similar quantity is  introduced  as a phenomenological parameter which describes
 polarizability of the core consisting of  outside nucleons.
 It should be noted that for this series of problems
 within the scope of the mean-field theory,   the self-consistent TFFS is
 similar to the HF-QRPA approach.

Recently,
quadrupole moments of the 
first $2^+$ state in
even lead and tin isotopes have been  found \cite{voitphysrev2012}, again
within the self-consistent TFFS. For this problem which is evidently
beyond the mean-field theory, the TFFS results are significantly
different from the QRPA ones.

In this paper, we review briefly the results of the cited
references on quadrupole moments of odd nuclei and of the $2^+$
states in even-even ones and add  some new calculations for odd unstable nuclei.
 In addition, we include here some results
of \cite{voitnsrt12} for the quadrupole moments of odd-odd nuclei
neighboring to the double magic ones.

\section{Brief calculation scheme}
The calculation scheme of the self-consistent TFFS based on the EDF
method by Fayans {\it et al.}  is described in detail in Ref.
\cite{BE2}. Here we write down only several formulas which are
necessary for understanding main ingredients of the approach. The
EDF method by Fayans {\it et al.} \cite{Fay} is a generalization for
superfluid finite systems of the original  Kohn--Sham EDF method
\cite{KSh}. In this method, the ground state energy of a nucleus is
considered as a functional of normal and anomalous densities, 
\beq
E_0=\int {\cal E}[\rho_n({\bf r}),\rho_p({\bf r}),\nu_n({\bf
r}),\nu_p({\bf r})] d^3r.\label{E0} 
\eeq

Within the TFFS, the static quadrupole moment $Q_{\lambda}$ of an
odd nucleus with the  odd nucleon in the state $\lambda$  can be
found in terms of the diagonal matrix element $ \langle\lambda|
V(\omega =0)|\lambda\rangle$  of the effective field $V$ in the
static external field $ V_0 = \sqrt{16\pi /5} r^2 Y_{20}$.
 In systems with pairing correlations, equation
for the effective field can be written in a compact form as 
\beq
\label{Vef_s}{\hat V}(\omega)={\hat e_q}V_0(\omega)+{\hat {\cal F}}
{\hat A}(\omega) {\hat V}(\omega), \eeq
 where all the terms  are
matrices. In the standard TFFS notation \cite{AB}, we have: \beq
{\hat V}=\left(\begin{array}{c}V
\\d_1\\d_2\end{array}\right)\,,\quad{\hat
V}_0=\left(\begin{array}{c}V_0
\\0\\0\end{array}\right)\,,
\label{Vs} \eeq

\beq {\hat {\cal F}}=\left(\begin{array}{ccc}
{\cal F} &{\cal F}^{\omega \xi}&{\cal F}^{\omega \xi}\\
{\cal F}^{\xi \omega }&{\cal F}^\xi  &{\cal F}^{\xi \omega }\\
{\cal F}^{\xi \omega }&{\cal F}^{\xi \omega }& {\cal F}^\xi \end{array}\right), \label{Fs} \eeq

\beq {\hat A}(\omega)=\left(\begin{array}{ccc} {\cal L}(\omega) &{\cal M}_1(\omega)
&{\cal M}_2(\omega)\\
 {\cal O}(\omega)&-{\cal N}_1(\omega) &{\cal N}_2(\omega)\\{\cal O}(-\omega)&-{\cal N}_1(-\omega) &
 {\cal N}_2(-\omega)
\end{array}\right)\,,
\label{As} \eeq where ${\cal L},\; {\cal M}_1$, and so on stand
for integrals over $\eps$ of the products of different
combinations of the Green function $G(\eps)$ and two Gor'kov
functios $F^{(1)}(\eps)$ and $F^{(2)}(\eps)$. They can be found in
\cite{AB}.

Isotopic indices in Eqs. (\ref{Vs}-\ref{As}) are omitted for
brevity. The explicit form of the above equations
 is written down for the case of the electric ($t$-even)
symmetry we deal with.
 In Eq. (\ref{Fs}), ${\cal F}$ is the usual LM amplitude,
\beq {\cal F}=\frac {\delta^2 {\cal E}}{\delta \rho^2}, \label{LM}\eeq
${\cal F}^{\xi}$ is the density-dependent effective pairing interaction,
\beq {\cal F}^{\xi}(\rho)=\frac {\delta^2 {\cal E}}{\delta \nu^2}, \label{EPI}\eeq
and the amplitudes ${\cal F}^{\omega \xi}={\cal F}^{\xi
\omega}$ stand for the mixed second derivatives,
\beq {\cal
F}^{\omega \xi}=\frac {\delta^2 {\cal E}}{\delta \rho \delta \nu}.
\label{LMxi}
 \eeq
In the case of volume pairing, one has ${\cal
F}^{\omega \xi}=0$, whereas for the case of surface pairing we deal the amplitude
${\cal F}^{\omega \xi}$ is non-zero and should be taken into account
when Eqs. (\ref{Vs}-\ref{As}) are solved. As the analysis of Ref. \cite{BE2} shows,
the component $V$ of the vector ${\hat V}$, as a rule, dominates. However, the fields
$d_1,d_2$  also contribute, and sometimes significantly, to the value of $Q_{\lambda}$.

In this article the TFFS equations are solved in the self-consistent
basis obtained within the EDF method with the functional DF3-a.
Thus, the same set of parameters  has been used to calculate the
single particle scheme and, according to Eqs.
(\ref{LM},\ref{EPI},\ref{LMxi}), the effective interactions in the
TFFS equations.  We consider the surface kind of pairing as
motivated by our previous research \cite{BE2}, see also {\it
ab initio} arguments in Ref. \cite{baldo2004}.

\begin{table}[ht!]
\caption{Quadrupole moments $Q\;$ $(e\;b)$ of odd-neutron nuclei in
the state $\lambda$.}

\begin{tabular}{lcccc }
\hline \hline nucleus  &$\lambda$  & $Q_{\rm exp}$
&\hspace*{1.5ex}$Q_{\rm th}$\hspace*{1.5ex}& $\delta Q$  \\

\hline
$^{39}$Ca &$1d_{3/2}$  &0.036(7)  &+0.040 &0.004 \\
          &            & 0.040(6) &       & 0.000 \\
$^{41}$Ca &$1f_{7/2}$ &-0.090(2)  &-0.078 &0.012\\
&                     &  -0.066(2)&       &-0.012  \\
&                     &-0.080(8)  &       &0.002\\
$^{85}$Kr  &$1g_{9/2}$  &  +0.443(3) &+0.507 &0.064  \\

$^{87}$Kr  &$2d_{5/2}$  &  -0.30(3)  &-0.355 &-0.06 \\

$^{87}$Sr  &$1g_{9/2}$  &  +0.33(2)  &+0.335& 0.01  \\

$^{89}$Sr  &$2d_{5/2}$  &  -0.271(9)  &-0.245  &-0.026 \\

$^{89}$Zr  &$1g_{9/2}$  &   +0.28(10) &+0.262  &-0.02 \\

$^{91}$Zr   &$2d_{5/2}$  &   -0.176(3) &-0.195 &-0.019 \\
            &            &   (-)0.257(13)       &     & 0.062 \\
            &            &   -0.206(10)         &    & 0.011 \\

$^{109}$Sn &$2d_{5/2}$ &+0.31(10)                  & +0.250 &-0.06 \\

$^{111}$Sn &$1g_{7/2}$ &+0.18(9)                   &+0.029   &-0.13\\

$^{115}$Sn &$1g_{7/2}^{\,*}$ &0.26(3)              &+0.377  &0.12\\

$^{119}$Sn &$2d_{3/2}^{\,*}$ &0.094(11) & -0.035   &-0.129  \\

           &                 &-0.065(5) &           & 0.030   \\

           &                 &-0.061(3) &           & 0.026   \\

$^{121}$Sn &$2d_{3/2}$ &-0.02(2)                 &+0.063 & 0.08  \\

$^{135}$Xe &$2d_{3/2}$ &    +0.214(7)               &+0.217& 0.003\\

$^{137}$Xe&  $2f_{7/2}$  &     -0.48(2)           &-0.376&0.10 \\

$^{137}$Ba&$2d_{3/2}$ &       +0.245(4)           &+0.254&0.009 \\

$^{139}$Ba& $2f_{7/2}$  &   -0.573(13)            & -0.445&0.128 \\

$^{141}$Nd &$2d_{3/2}$ &    +0.32(13)             &+0.289 &-0.03 \\

$^{143}$Nd&  $2f_{7/2}$  &  -0.61(2)              &-0.518 &0.09 \\

$^{143}$Sm &$2d_{3/2}$ &     +0.4(2)              &+ 0.296 &0.1\\

$^{145}$Sm&  $2f_{7/2}$  &  -0.60(7)              &-0.537 &0.06 \\

$^{197}$Pb &$3p_{3/2}$ &-0.08(17)                &+0.195 &0.27  \\

$^{199}$Pb &$3p_{3/2}$ &+0.08(9)                 &+0.272   & 0.19\\

$^{201}$Pb &$2f_{5/2}$ &-0.01(4)                 &+0.137  & 0.15\\

$^{203}$Pb &$2f_{5/2}$ &+0.10(5)                  &+0.284  & 0.18\\

$^{205}$Pb &$2f_{5/2}$ &+0.23(4)                 &+0.336   &0.09  \\

$^{209}$Pb &$2g_{9/2}$ &-0.3(2)                  &-0.264   &0.1\\

$^{211}$Pb &$2g_{9/2}$ &+0.09(6)                 &-0.283   &-0.37\\

\hline\hline

$^{113}$Sn &$1h_{11/2}^{\,*}$ &0.41(4)          &-0.776 & -0.37 \\
           &                  & 0.48(5)       &          & -0.30  \\

$^{115}$Sn &$1h_{11/2}^{\,*}$ &0.38(6)           &-0.703   &-0.32 \\

$^{117}$Sn &$1h_{11/2}^{\,*}$ &-0.42(5)          &-0.593& -0.17  \\

$^{119}$Sn &$1h_{11/2}^{\,*}$ &0.21(2)           &-0.469&  -0.25 \\

$^{121}$Sn &$1h_{11/2}^{\,*}$ &-0.14(3)          &-0.293  &-0.15 \\

$^{123}$Sn &$1h_{11/2}$ &+0.03(4)                &-0.123   &-0.15 \\

$^{125}$Sn &$1h_{11/2}$ &+0.1(2)                 &+0.039   & -0.1  \\

$^{135}$Xe &$1h_{11/2}^{\,*}$ &   +0.62(2)         &+0.504&0.12 \\

$^{137}$Ba &$1h_{11/2}^{\,*}$ &  +0.78(9)            &+0.588&-0.19 \\

$^{147}$Gd& $1i_{13/2}^{\,*}$ &  -0.73(7)          & -0.791 &-0.06 \\

$^{191}$Pb &$1i_{13/2}^{\,*}$ &+0.085(5)          &+0.0004 &-0.085   \\

$^{193}$Pb &$1i_{13/2}^{\,*}$ &+0.195(10)         &+0.335  &0.140  \\

$^{195}$Pb &$1i_{13/2}^{\,*}$ &+0.306(15)         &+0.689  &0.383  \\

$^{197}$Pb  &$1i_{13/2}^{\,*}$ &+0.38(2)          &+0.980  &0.60   \\

 $^{205}$Pb &$1i_{13/2}^{\,*}$& 0.30(5)           &+0.665  &0.37  \\

\hline \hline
\end{tabular}
\label{tab:Q_n}
\end{table}

\section{Quadrupole moments of odd semi- and near-magic nuclei}
The final expression for the quadrupole moment of an odd nucleus is
as follows \cite{AB,solov}:

 \beq\label{Qlam}
 Q^{p,n}_{\lambda} =  (u^2_{\lambda}-v^2_{\lambda}) V^{p,n}_{\lambda},
 \eeq
 where $u_{\lambda}$, $v_{\lambda}$ are the Bogolyubov coefficients and
 \beq\label{Vlam}
 V_{\lambda} =  -\frac{2j-1}{2j+2} \int  V(r) R_{nlj}^2(r)
 r^{2}dr.
 \eeq
For odd neighbors of a magic nucleus the ``Bogolyubov'' factor in
(\ref{Qlam}) reduces to 1 for a particle state and to $-1$ for a
hole one, see also \cite{BM}. If the odd nucleon belongs to the
superfluid component,  the factor $(u^2_{\lambda}-v^2_{\lambda})$ in
Eq. (\ref{Qlam}) becomes non-trivial. It changes permanently
depending on the state $\lambda$ and the nucleus under
consideration.  This factor determines the sign of the quadrupole
moment. It depends essentially on values of the single-particle
basis energies $\eps_{\lambda}$ reckoned from the chemical potential
$\mu$ as we have
\beq(u^2_{\lambda}-v^2_{\lambda})=(\eps_{\lambda}-\mu)/E_{\lambda},\eeq
$E_{\lambda}=\sqrt{\eps_{\lambda}^2+\Delta_{\lambda}^2}$. Keeping in
mind such sensitivity, we found this quantity for a given odd
nucleus $(Z,N+1)$ or $(Z+1,N)$,  $N,Z$ even, with taking into
account the blocking effect in the pairing problem \cite{solov}
putting the odd nucleon to the state $\lambda$ under consideration.
For the $V_{\lambda}$ value in Eq. (\ref{Qlam}) we used the half-sum
of these values in two neighboring even nuclei.

The results of the calculations are presented in Tables 1 and 2
which contain odd-neutron and odd-proton nuclei respectively with
known experimental quadrupole moments (our predictions for odd
nuclei with unknown quadrupole moments see in \cite{BE2,QM}).
One can see that the theoretical sign of the quadrupole
moment is correct in all cases when the sign of the experimental
moment is known. This permits to use our predictions to determine
the sign when it is unknown. Several rather strong disagreements
with the experimental data for high-j levels $1h_{11/2}$ in Sn
isotopes and $1i_{13/2}$ in Pb isotopes originate from their too
distant positions from the Fermi level, see \cite{QM}, where
it was found that the $Q$ values depend strongly on the
single-particle level structure. It follows mainly from Eq. (11).

\begin{table}[ht!]
\caption{Quadrupole moments $Q\;$(b) of odd-proton nuclei in the
state $\lambda$.}

\begin{tabular}{l c c c  c}
\hline \hline nucl.  &$\lambda$  & $Q_{\rm exp}$ &\hspace*{1.ex}
$Q_{\rm
th}$\hspace*{1.ex}&$\delta Q$\\

\hline
$^{39}$K & $1d_{3/2}$&0.0585(6)     &0.069      &0.010  \\

$^{41}$Sc & $1f_{7/2}$&-0.156(3)    &-0.139     & 0.017 \\
&                     &0.120(6)    &            &-0.019\\
&                     &0.168(8)  &              & 0.029\\

$^{87}$Rb & $2p_{3/2}$ & +0.134(1)    &+0.132  &-0.002\\
&                      &  +0.138(1)   &        &-0.006\\

$^{105}$In & $1g_{9/2}$& +0.83(5) &   +0.833    & 0.00 \\

$^{107}$In & $1g_{9/2}$& +0.81(5) &+0.976       &0.17   \\

$^{109}$In & $1g_{9/2}$& +0.84(3) &+1.113       & 0.27 \\

$^{111}$In & $1g_{9/2}$& +0.80(2) &+1.165       &0.36 \\

$^{113}$In & $1g_{9/2}$& +0.80(4) &+1.117       &0.32\\

$^{115}$In & $1g_{9/2}$& +0.81(5) & +1.034      &0.22  \\

&                       & 0.58(9)&              &0.45\\

$^{117}$In & $1g_{9/2}$& +0.829(10)& +0.965     & 0.136  \\

$^{119}$In & $1g_{9/2}$& +0.854(7) &+0.909      & 0.055\\

$^{121}$In & $1g_{9/2}$& +0.814(11) &+0.833     &0.019 \\

$^{123}$In & $1g_{9/2}$& +0.757(9)  &+0.743     &-0.014   \\

$^{125}$In & $1g_{9/2}$& +0.71(4)   &+0.663     &-0.05 \\

$^{127}$In & $1g_{9/2}$& +0.59(3)   &+0.550    &-0.04  \\

$^{115}$Sb & $2d_{5/2}$& -0.36(6)   &-0.882   & -0.52       \\

$^{119}$Sb & $2d_{5/2}$& -0.37(6)   &-0.766   &-0.40 \\

$^{121}$Sb & $2d_{5/2}$& -0.36(4)   & -0.721 & -0.36 \\
&                      & -0.45(3)   &             & -0.27\\

$^{123}$Sb  & $1g_{7/2}$& -0.49(5)  &-0.739 & -0.25    \\

$^{137}$Cs  & $1g_{7/2}$&+0.051(1)    &-0.031  &-0.080 \\

$^{139}$La &  $1g_{7/2}$& +0.20(1)    &+0.103  &-0.10 \\

$^{141}$Pr &   $2d_{5/2}$& -0.077(6)  &-0.120  &-0.043 \\
&                        &  -0.059(4)  &       &-0.061\\

$^{145}$Eu &   $2d_{5/2}$&   +0.29(2)  &+0.156 &-0.13 \\

$^{205}$Tl &   $3d_{3/2}^*$ &+0.74(15)   & +0.227 &-0.51 \\

$^{203}$Bi  & $1h_{9/2}$ & -0.93(7) & -1.323 &-0.39 \\
            &           &  -0.68(6) &        & -0.64 \\
$^{205}$Bi  & $1h_{9/2}$& -0.81(3)  & -0.945  &-0.14 \\
            &           & -0.59(4)  &         &-0.36 \\
$^{207}$Bi  & $1h_{9/2}$& -0.76(2)  & -0.454  &0.31 \\
            &           & -0.55(4)  &        &0.10 \\
            &           &-0.60(11)  &        &0.15 \\

$^{209}$Bi  & $1h_{9/2}$& -0.516(15)   & -0.342 & 0.18  \\
&                       & -0.37(3)     &        & 0.03\\
&                       &  -0.55(1) &           & 0.21\\
&                       &  -0.77(1) &           &  0.43 \\
&                       &  -0.40(5) &           &  0.06 \\
&                       &  -0.39(3) &           &  0.05 \\

$^{213}$Bi  & $1h_{9/2}$& -0.83(5)  &-0.508   & 0.32 \\
            &           & -0.60(5)  &         & 0.09 \\

\hline \hline
\end{tabular}
\label{tab:Q_p}
\end{table}

To evaluate the agreement with experiment quantitatively, we
calculate the mean theory-experiment difference
 \beq
\sqrt{\overline{(\delta Q)^2_{\rm rms}}}  =  \sqrt{\frac 1 {\cal N}
\sum_{i} \left(Q^{\rm th}_i- Q^{\rm exp}_i\right)^2},\label{rms}
\eeq
with obvious notation. On average, the agreement,  can be considered
as reasonable. For 42 quadrupole moments of odd-neutron nuclei, the
average disagreement between theory and experiment is not so small,
$ \sqrt{\overline{(\delta Q)^2_{\rm rms}}} = 0.189\;$e b. However,
it is concentrated mainly in 15 intruder states for which we have  $
\sqrt{\overline{(\delta Q)^2_{\rm rms}}}[\rm intruder] = 0.269\;$e
b. For the rest of
 27 ``normal'' states, the disagreement is rather moderate
$ \sqrt{\overline{(\delta Q)^2_{\rm rms}}}[\rm normal] =  0.125\;$e
b. For protons, agreement is worse. The rms deviation is $
\sqrt{\overline{(\delta Q)^2_{\rm rms}}} = 0.254\;$e b. The main
contribution to this deviation comes from In and Sb isotopes, odd
neighbors of even tin nuclei. It is the result of too strong
quadrupole field $V_{n,p}(r)$  for the DF3-a functional \cite{BE2}.
For neutrons, this drawback is partially hidden with multiplying by
the Bogolyubov factor, but for protons it appears to the full
extent. For more detailed discussion, see \cite{QM}.

For odd-neutron neighbors of even $N {=} 50$ isotones, the
 proton-subsystem is superfluid and the neutron Bogolyubov factor in Eq. (\ref{Qlam})
 is $\pm1$. In this case, agreement with the data is almost perfect,
$ \sqrt{\overline{(\delta Q)^2_{\rm rms}}} {=} 0.041\;$e b. The
situation is similar for odd-neutron neighbors of even isotones with
$N{=}82$. Again, agreement with the data is rather good, $
\sqrt{\overline{(\delta Q)^2_{\rm rms}}} {=} 0.093\;$e b.

 For the major part of nuclei in Tables 1 and 2, neighboring
to double-magic ones, let us call them ``near-magic'', the quality
of agreement is rather good. Therefore, we hope to predict
reasonably the quadrupole moment  values for such nuclei including
strongly proton- or neutron-rich ones. These predictions are
presented in Table 3.

\begin{table}[th]
\label{tablqoddn} \caption{Predictions for quadrupole moments $Q\;$
$(e\;b)$ of odd near-magic nuclei.}
\begin{center}
\begin{tabular}{l c c c c c }
\hline \hline \noalign{\smallskip}
 nucl.  & $J^{\rm \pi}$
&\hspace*{1.ex} $T_{\rm 1/2}$\hspace*{1.ex} &\hspace*{1.ex} $Q_{\rm theor}$ &\hspace*{1.ex}$Q_{\rm exp}$\\
\noalign{\smallskip}
\hline
\noalign{\smallskip}
${^{55}_{28}}$Ni$_{27}$ & 7/2$^{+}$    & 204.7 ms    & -0.26  & --\\
${^{57}_{28}}$Ni$_{29}$ & 3/2$^{-}$    & 35.6 h      & -0.17  & --\\
${^{77}_{28}}$Ni$_{49}$ & (9/2)$^{+}$  & 128 ms      &  0.20  & --\\
${^{79}_{29}}$Ni$_{50}$ & (5/2$^{+}$)  & 635 ns      & -0.12  & --\\
${^{101}_{50}}$Sn$_{51}$ & (5/2)$^{+}$ & 1.7 s       & -0.21  & --\\
${^{131}_{50}}$Sn$_{81}$ & (3/2$^{+}$) & 56 s        &  0.10  & -0.04(8)\\
${^{133}_{50}}$Sn$_{83}$ & 7/2$^{-}$   & 1.46 s      & -0.17  & --\\
${^{207}_{82}}$Pb$_{125}$ & (1/2)$^{-}$& stable      &  0     & --\\
${^{55}_{27}}$Co$_{28}$ & 7/2$^{-}$    & 17.53 h     &  0.31  & --\\
${^{57}_{29}}$Cu$_{28}$ & 3/2$^{-}$    & 196.3 ms    & -0.20  & --\\
${^{79}_{29}}$Cu$_{50}$ & (3/2$^{-}$)  & 188 ms      & -0.13  & --\\
${^{99}_{49}}$In$_{50}$ & (9/2)$^{+}$  & 3 s         & -0.35  & --\\
${^{131}_{49}}$In$_{82}$ & (9/2)$^{+}$ & 0.28 s      &  0.28  & --\\
${^{133}_{51}}$Sb$_{82}$ & (7/2)$^{+}$ & 2.34 m      & -0.23  & --\\
${^{207}_{81}}$Tl$_{126}$ & 1/2$^{+}$  & 4.77 m      &  0     & --\\
\\
\hline \hline
\end{tabular}
\end{center}
\end{table}

 As it was mentioned  in Introduction,
 the core polarizability  by  the
quadrupole external field is characterized  directly by the
effective quadrupole charges, which are defined  naturally within
TFFS as  $e_{\rm eff}^{p,n}=V^{p,n}_{\lambda}/(V_0^p)_{\lambda}$
\cite{kaevyadfiz1965}. In  Tables 1 and 2, there are  only two
nucleus, $^{209}$Bi and $^{209}$Pb, with a double-magic core. In
this case, the polarizability is relatively  moderate, $e^{p}_{\rm
eff}=1.4$, $e^n_{eff}= 0.9$. In nuclei with unfilled neutron shell,
it becomes much stronger, $e_{\rm eff}\simeq 3\div 6$  \cite{BE2}.
The reason is rather obvious. Indeed, for the case of positive
parity field $V_0$, virtual transitions inside the unfilled shell
begin to contribute in such nuclei and small energy
denominators appear in the propagator ${\cal L}^n$, Eq. (5),
 playing the main role in Eq. (\ref{Vef_s}) for the problem under consideration.
This enhances the neutron response to the field $V_0$ and, via the
strong LM neutron-proton interaction amplitude ${\cal F}^{np}$, the
proton response as well. The results in Table 2 for the chain
$^{203,205,209}$Bi show how the polarizability grows with increase
of the number of neutron holes. Keeping in mind this physics, one
can represent the effective charges as  $e^{p}_{\rm eff} = 1 +
e^{p}_{\rm pol}, e^{n}_{\rm eff} = e^{n}_{\rm pol}$ where
$e^{p,n}_{\rm pol}$ is the pure polarizability charge. To separate
contributions of the unfilled shells and core nucleons explicitly,
one can divide the Hilbert space of the QRPA equations (\ref{Vef_s})
to the ``valent'' and subsidiary ones and carry out the
corresponding renormalization procedure  \cite{kaev1969}.

\section{ Quadrupole moments of the first 2$^+$ states in Sn and Pb isotopes }
Account for the phonon coupling (PC) is the direct way to
generalize the standard nuclear theory. As a rule, the so-called
$g^2$ approximation is used where $g$ is the phonon creation
amplitude. However, almost all of these generalizations
did not take into account \textit{all} the $g^2$ terms,
limiting themselves with the mass operator pole diagrams only, see
the first diagram in Fig. 1, where diagrams for the mass operator
are displayed. The second diagram represents the sum of all $g^2$
 non-pole diagrams  usually called the  tadpole.
  The problem of consistent consideration of all  $g^2$
 terms including tadpoles was analyzed firstly in the  article by Khodel  \cite{khodel1976}.
The method developed was applied  to magic nuclei, mainly for ground
state nuclear characteristics, within  the self-consistent TFFS
\cite{khodelsap}. It was found
 that, as a rule, the tadpole contributions in magic nuclei
 are noticeable and are often of opposite sign as compared with those of the
 pole terms. The first attempts to include  phonon  tadpole effects for nuclei with pairing
 were  recently made in Refs. \cite{kaevsap},
 \cite{kaevavevoit2011} and \cite{voitphysrev2012}.

\begin{figure}
\resizebox{1.\columnwidth}{!}{%
\includegraphics{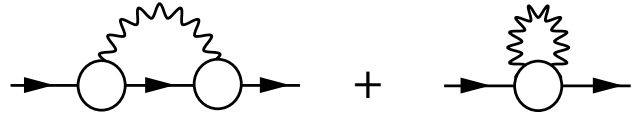}}
\caption{$g^2$ order corrections to the mass operator
in magic nuclei. The circles with one wavy line in the first term
are the phonon creation amplitudes $g$. The second term is the phonon
tadpole.}
\end{figure}

\vspace{3mm} According to Ref. \cite{khodel1976}, in the $g^2$
approximation, the matrix element $M_{LL}$ for a static moment of
the excited state (phonon) with the orbital angular moments $L$ in a
static external field $V^{0}$ is determined in terms of the change
of  the one-particle Green function (GF) in the field of this
phonon:
\begin{equation}
\label{matrixelem} M_{LL} = \int V^{0}(\textbf{r})\delta^{(2)}_{LL}
G(\textbf{r},\textbf{r},\varepsilon )d\textbf{r}
\frac{d\varepsilon}{2 \pi  \imath} ,
\end{equation}
 \bea \label{deltaLL} \delta^{(2)}_{L L} G =\delta_{L}(G g_{L}
G) = G(\varepsilon) g_{L}
G(\varepsilon +  \omega_{L}) g_{L} G(\varepsilon) \\
 \nonumber +G(
\varepsilon) g_{L} G( \varepsilon - \omega_{L}) g_{L} G(
\varepsilon)+ G( \varepsilon) \delta_L g_{L} G( \varepsilon), \eea
where $g_L$ is the amplitude for the production of the L phonon with
the  energy $\omega_L$ and $ \delta_L g_{L}$ is  the variation of
$g_L$ in the  field of other $L$ phonon. This quantity is the main
part of  the phonon tadpole in Fig. 1.  After some transformations
of these expressions one can obtain
\begin{equation}
\label{matrixelem4} M_{L L} = V^0 G g_L G g_{L} G + V^0 A \delta_L
g_{L}.
\end{equation}

It is convenient to transform this expression in such a way that the
effective field $V$, Eq. (\ref{Vef_s}), appears instead of the
external field $V^{0}$. After regrouping  terms in Eq.
(\ref{matrixelem4}) and in the integral  equation for
   $\delta_L g_L $, for  details, see Refs.
\cite{khodelsap,kaevavevoit2011,voitphysrev2012}, we obtain the
ultimate expression,
\begin{equation}
\label{matrixelem2} M_{L L}= VGg_LGg_LG +V A \delta_L {\cal F} A
g_{L},
\end{equation}
which is illustrated in Fig. 2.
 It contains now the effective field $V$, Eq. (\ref{Vef_s}), instead of  $V^{0}$
  and the quantity $\delta_L {\cal F}$ in the second term which
 denotes  the variation of the effective ph interaction  ${\cal F}$ in
 the field of the $L$ phonon. For the density dependent TFFS
 effective  interaction ${\cal F}(\rho)$, the following ansatz can
 be readily obtained \cite{khodel1976,khodelsap}:
\begin{equation}
\label{deltaLF} \delta_L {\cal F}({\bf r})= \frac{\partial {\cal
F}}{\partial \rho}  \rho_L^{\rm tr}(r) Y_{LM}(\bf n),
\end{equation}
 where $\rho_L^{\rm tr}=Ag_L$ is the transition
density for the $L$ phonon excitation. The first term of Eq. (16)
coincides with the result of Refs. \cite{speth1970,Br1970} while the
second one, with the $ \delta_L {\cal F} $ quantity, is a
generalization  to take into account all the $g^2$ terms.

\begin{figure}
\resizebox{1.\columnwidth}{!}{%
\includegraphics{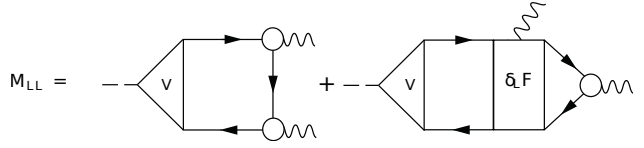}}
\caption{Matrix element  $M_{L L}$ in the form of Eq. (16).}
\end{figure}

All the above equations can be readily modified for such processes
as the transition between the excited states $L$ and $L'$ in the
external field $V^0(\omega = \omega_{L'}- \omega_L)$ or the
excitation of the two-phonon state $L+L'$ in the external field
$V^0(\omega = \omega_{L'}+\omega_L)$. The static moment case corresponds
to $\omega = 0, \omega_{L'}= \omega_L $.

This approach for magic nuclei has been generalized for non-magic
ones in \cite{kaevvoit2009,voitphysrev2012}. Then eight matrix
elements instead of one in Fig. 2 should be considered, two of them
are shown  in Fig. 3.

\begin{figure}
\resizebox{1.\columnwidth}{!}{%
\includegraphics{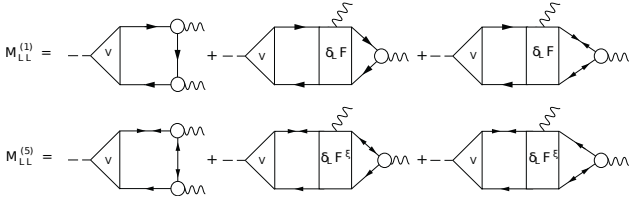}}
\caption{Matrix elements for  $M^{(1)}_{L L}$ and $M^{(5)}_{L L}$
for non-magic nuclei.}
\end{figure}

It is necessary to compare this  expression with the  QRPA approach.
Here we mean  the usual scheme \cite{ponomarev1998} which uses the QRPA wave functions for
the matrix element between two excited states.
In Ref. \cite{ponomarev1998} the expression for
the $B(E2)$ quantity has been derived using  the
bare external  field $V^0$ and
  the QRPA wave
functions without the pp and hh-channels. The analytical expression
for the sum of the eight above-mentioned matrix elements consists of
two parts. The first part coincides with the corresponding formula
in \cite{ponomarev1998,solov} with one important correction, which
is the first generalization of the QRPA approach. Namely, instead of
the external field $V^0$, which does not depend on the frequency,
 the effective field $V$ appears, which depends in general on the
frequency $\omega = \omega_L \pm \omega_L^\prime $. The second  part
of the sum is new and describes the contribution of the ground state
correlations (GSC), the so-called backward-going diagrams, to the
first diagrams of Fig. 3 with the integrals of three GF's
(``triangle''). This is the second generalization. We calculate the
contribution of such correlations separately.
 The terms with $\delta_L {\cal F}$ and
$\delta_L {\cal F}^\xi $, Fig. 3, are the third generalization of
the QRPA approach. Note that these terms are also absent in Refs.
\cite{Br1970,speth1970,vdovin1,vdovin2,broglia1972}.
   The main difference of our approach from the calculations  in Refs.
\cite{Br1970,broglia1972,vdovin1,vdovin2} is
the self-consistency on the (Q)RPA level and absence of any
phenomenological or fitted parameters.

\begin{table}[t]
\caption{Quadrupole moments $Q\;$($e\;b)$ of the first 2$^+$ states in
Sn and Pb isotopes.}

\begin{tabular}{l c c c c}
\hline \hline \noalign{\smallskip}
nucl.  & \hspace*{1.ex}$Q_{\rm theor}$\hspace*{1.ex} &\hspace*{1.ex}
$Q_{\rm exp}$ \cite{stone}&\hspace*{1.ex}$Q (GSC=0)$&\hspace*{1.ex}$Q_{\rm QRPA}$\\
\noalign{\smallskip}
\hline
\noalign{\smallskip}
$^{100}$Sn & 0.04 &--  & 0.05 & 0.017\\
$^{102}$Sn &-0.07 &--  & -0.02 & -0.001\\
$^{104}$Sn &-0.22 &--  & -0.08 & -0.001\\
$^{106}$Sn &-0.34 &--  &-0.13 & -0.002\\
$^{108}$Sn &-0.39 &--  &-0.14 & -0.002\\
$^{110}$Sn &-0.50 &--  &-0.17 & -0.003\\
$^{112}$Sn &-0.45 &-0.03(11)  &-0.15 & -0.003\\
$^{114}$Sn &-0.28 &0.32(3),  &-0.09 & -0.004\\
 &  & 0.36(4)\\
$^{116}$Sn &-0.12 &-0.17(4),  & -0.03 & -0.003\\
 & & +0.08(8)\\
$^{118}$Sn &-0.01 &-0.05(14)  & 0.01 & -0.003\\
$^{120}$Sn & 0.04 &+0.022(10),  &0.03 & -0.003\\
 & & -0.05(10)\\
$^{122}$Sn & 0.01 &-0.28 $<Q$  &0.02 & -0.003\\
 & &$Q<$+0.14&\\
$^{124}$Sn &-0.07 &0.0(2)  & -0.01 & -0.003\\
$^{126}$Sn &-0.13 &--  &-0.04 & -0.002\\
$^{128}$Sn &-0.14 &--  & -0.05 & -0.002\\
$^{130}$Sn &-0.07 &--  & -0.03 & -0.001\\
$^{132}$Sn & 0.04 &--  & 0.05 & 0.015\\
$^{134}$Sn &-0.01 &--  & 0.00 & -0.001\\
\noalign{\smallskip}
\hline
\noalign{\smallskip}
$^{190}$Pb &-0.92 &--  & -0.30 & -0.008\\
$^{192}$Pb &-1.15 &--  & -0.38 & -0.008\\
$^{194}$Pb &-1.31 &--  & -0.44 & -0.008\\
$^{196}$Pb &-1.26 &--  & -0.42 & -0.008\\
$^{198}$Pb &-1.05 &--  & -0.35 & -0.008\\
$^{200}$Pb &-0.52 &--  & -0.17 & -0.006\\
$^{202}$Pb &-0.15 &--  & -0.03 & -0.005\\
$^{204}$Pb & 0.10 &+0.23(9)  & 0.06 & -0.003\\
$^{206}$Pb & 0.09 &+0.05(9)  & 0.06 & -0.002\\
$^{208}$Pb & 0.05 &-0.7(3)   & 0.07 & 0.043\\
\\
\hline \hline
\end{tabular}
\end{table}

\begin{figure}
\resizebox{1.\columnwidth}{!}{%
\includegraphics{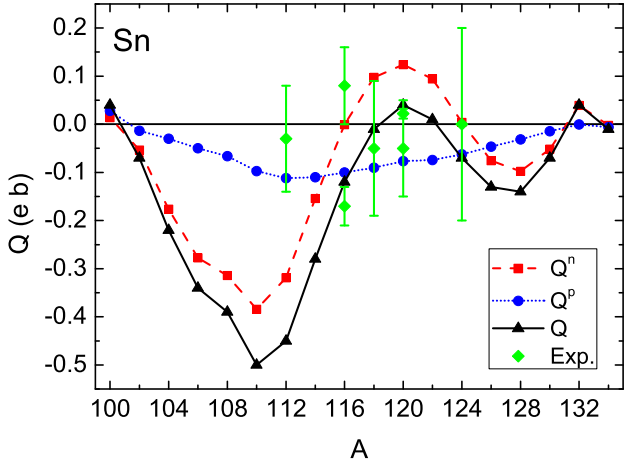}}
\caption{Quadrupole moments of the first 2$+$ excited states in even
Sn isotopes.}
\end{figure}

  We calculated the quadrupole moments of the first 2$^+$ states in
  the tin and lead isotopes
  in the $\lambda$-representation with
  self-consistent single-particle wave functions $\phi_{\lambda}$ obtained within the EDF method
  of Ref. \cite{Fay} with the functional DF3-a \cite{Tol-Sap1}.
  A spherical box
  of the radius $R{=}16\;$fm is used to simulate the single-particle continuum. We examined the
  dependence of the  results on the cut-off energy $E_{\rm max}$
  and have found that the value of $E_{\rm max}{=}100\;$MeV ensures 1\% accuracy.  To calculate
  the quantities V and $g_L$, the  results of Ref. \cite{BE2} have been used where
  all the calculations were performed in the coordinate representation using the same self-consistent
  DF3-a basis as in the present calculation of the matrix element  $M_{LL}$.
 Thus, the  single-particle continuum is taken into account adequately in the present
 calculations.
 The contribution to the Q values of the term with  $\delta \cal F$ in Fig. 3,
 Eq. (\ref{deltaLF}), turned out to be rather small,
 ($-(0.01\div 0.03)\;$e b).
 However, there are cases where these corrections are comparable with the total $Q(2^{+}_{1})$ value
 when the proton and neutron
  values almost compensate each other, e.g. in $^{118}$Sn and $^{122}$Sn nuclei.
The term with $\delta_L{\cal F}^\xi$ contains the anomalous analogs
of the corresponding quantities in Eq. (\ref{deltaLF}).

  The results are given in Table 4 and Fig. 4. Except for  $^{112}$Sn
  and $^{208}$Pb nuclei,
  we obtained a reasonable agreement with experimental data
  \cite{stone}.
The contribution of the GSC term
 turned out to be large.
Often it is more than $50\div 60$\% of all  triangle contributions
(column Q(GSC=0)). The usual QRPA (GSC=0 and $V = V^0$), see the
last column in Table 4, results in considerably less Q values.

\section{Quadrupole moments of  odd-odd near magic nuclei}
As the odd-odd nuclei are more complicated objects than the odd
ones, we consider here only the near-magic odd-odd nuclei.
Within the above-described self-consistent EDF approach, we
calculated  the ground state quadrupole moments of odd-odd
near-magic nuclei with the use
of the approximation disregarding the 
interaction between two odd particles.
This simple approximation
can be checked
in a pure phenomenological way \cite{kaevyadfiz,voitnsrt12} and
it turned out reasonable.
Within this approximation, the problem is reduced to
calculations of quadrupole moments of corresponding odd nuclei.

Indeed, if we neglect the interaction between two quasi-particles,
the quadrupole moment of the odd-odd nucleus with the spin $I$ is as
follows:

\beq\label{qu} Q_I=<II \mid V^p + V^n \mid II>, 
\eeq 
where $\Psi_{II}
= \Sigma \varphi_1 \varphi_2 <j_1m_1 j_2m_2\mid II>,$ for the
particle-particle case.
Here  $\varphi_1$ is the
single-particle wave function with the quantum numbers $1 \equiv
\lambda_1 \equiv (n_1,j_1,l_1,m_1)$.
Then the expression for ground state quadrupole moment of the odd-odd
near-magic nucleus  has the form:
 \bea\label{oddodd}
 Q_I = (2I+1)
\left(\begin{array}{ccc}
{I} &{2} &{I}\\
{I} &{0} &{-I} \end{array}\right)
(-1)^{j_p+j_n+I+2} \times\nonumber\\
\times \left[\left\lbrace\ \begin{array}{ccc}
{j_p} &{I} &{j_n}\\
{I} &{j_p} &{2}
\end{array} \right\rbrace c^{-1}_{j_{p}} Q^{p} 
+ \left\lbrace \begin{array}{ccc}
j_n & I & j_p\\
I & j_n & 2
\end{array}\right\rbrace  c^{-1}_{j_{n}} Q^{n}  \right] ,
\eea where 3j-symbol $c_{j}=2j(2j-1)^{1/2}\left[(2j+3)(2j+2)\right.
\times $ \\$\left. \times(2j+1)2j \right]^{-1/2}$, $Q^p$ and $Q^n$
are the quadrupole moments of corresponding odd nuclei. Similar
formulae  can be easily obtained for the hole-hole, particle-hole
and hole-particle  cases. The details can be found in the
Poster article by Voitenkov {\it et al.} \cite{voitnsrt12} at the given 
conference.

The results of the calculations are presented in Table 5. In column
Q${\rm _{eff}}$ we show  the results for the odd-odd nuclei
under consideration obtained with the effective charges
$e{^p_{eff}}=2$, $e{^n_{eff}}=1$ in order to compare them   with this
well-known phenomenological description. We see that there are only
three experimental Q values for the long-living nuclei and our
approach describes them rather satisfactory.
 Other nuclei are short-living ones and our results give  reliable predictions for their Q values.

\begin{table}[h]
\label{selfodd-odd}
\caption{Quadrupole moments \textit{Q} (e b) of odd-odd near-magic nuclei.}
\begin{center}
\begin{tabular}{l c c c c c }
\hline \hline \noalign{\smallskip}
 nucl.  & $J^{\rm \pi}$
&\hspace*{1.ex} $T_{\rm 1/2}$\hspace*{1.ex} &\hspace*{1.ex}$Q_{\rm eff}
$\hspace*{1.ex} &\hspace*{1.ex} $Q_{\rm theor}$ &\hspace*{1.ex}$Q_{\rm exp}$\\
\noalign{\smallskip}
\hline
\noalign{\smallskip}
${^{54}_{27}}$Co$_{27}$ & 0$^{+}$      & 193.28 ms  &  --   & --     & --\\
${^{56}_{27}}$Co$_{29}$ & 4$^{+}$      & 77.236 d   &  0.19 &  0.30  & +0.25(9)\\
${^{56}_{29}}$Cu$_{27}$ & (4$^{+}$)    & 93 ms      &  0.14 &  0.28  & --\\
${^{58}_{29}}$Cu$_{29}$ &  1$^{+}$     & 3.204 s    &  0.09 &  0.15  & --\\
${^{78}_{29}}$Cu$_{49}$ & (3$^{-}$)    & 637 s      & -0.18 & -0.21  & --\\
                        & (4$^{-}$)    &            &  4$\times 10^{-5}$ &  -0.03  & --\\
${^{100}_{49}}$In$_{51}$& (6$^{+}$)    & 5.9 s      &  0.24 &  0.21  & --\\
${^{130}_{49}}$In$_{81}$&  1$^{-}$     & 0.29 s       & -0.08 & -0.07  & --\\
${^{132}_{49}}$In$_{83}$& (7$^{-}$)    & 0.207 s    & -0.40 & -0.29  & --\\
${^{132}_{51}}$Sb$_{81}$& (4)$^{+}$    & 2.79 m     & -0.30 & -0.22  & --\\
${^{134}_{51}}$Sb$_{83}$& (0$^{-}$)    & 0.78 s     &  --   & --     & --\\
${^{206}_{81}}$Tl$_{125}$& 0$^{-}$     & --         & --    & --     & --\\
${^{208}_{81}}$Tl$_{127}$& 5$^{+}$     & 3.053 m    & -0.30 & -0.27  & --\\
${^{208}_{83}}$Bi$_{125}$& 5$^{+}$     & 3.68E+5 y  & -0.51 & -0.35  & -0.64(6)\\
${^{210}_{83}}$Bi$_{127}$& 1$^{-}$     & 5.012 d    &  0.21 &  0.16  & +0.136(1)\\
\\
\hline \hline
\end{tabular}
\end{center}
\end{table}

\section{Conclusion}

 Quadrupole moments  of odd neighbors of
semi-magic lead and tin isotopes and $N=50,N=82$ isotones are
calculated within the self-consistent TFFS  based on the Energy
Density Functional by Fayans {\it et al.} with the  DF3-a parameters 
fixed previously. The same approach has been used to calculate
quadrupole moments of the first 2$^+$ state in tin and lead isotopes as
well as the moments of near-magic odd-odd nuclei.

For the quadrupole moments  of odd and odd-odd near magic nuclei a good agreement with 
the experiment has been obtained.For the case of semi-magic nuclei a reasonable agreement with  experiment for the  quadrupole moments has been obtained for the most part of nuclei considered. In
this case when the odd particle belongs to the superfluid subsystem, the Bogolyubov factor
$(u^2_{\lambda}-v^2_{\lambda})=(\eps_{\lambda}-\mu)/E_{\lambda}$
comes to the quadrupole moment value, in addition to the matrix
element of the effective field $V_{\lambda}$.  This factor makes the
quadrupole moment value very sensitive to calculation accuracy of 
the  single-particle energy $\eps_{\lambda}$ of the state under
consideration, especially near the Fermi surface as the quantity
$Q_{\lambda}$ vanishes at $\eps_{\lambda}=\mu$.  For such a
situation, influence of the coupling of single-particle degrees of
freedom with phonons, see \cite{Tol-Sap,kaevavevoit2011,QM}, should
be especially important.

For the quadrupole moments of the first $2^+$ states, we have obtained a
noticeable difference from the traditional QRPA approach. In
particular, new terms with $\delta_{L}{\cal F}$ and $\delta_{L}{\cal
F}^{\xi}$ appear, which  contain the density derivatives of both the
ph and pp effective interactions. In the problem under
consideration, their contribution turned out  to be
rather small, as a rule. However, for consistency, these terms should be
included. Except for the $^{112}$Sn and $^{208}$Pb cases, a
reasonable  agreement has been obtained with the experiment
available. Using the self-consistent method which contains no  newly
adjusted parameters we have also predicted the values of quadrupole
moments of the first 2$^+$ states in several unstable lead and tin
isotopes including the $^{100}$Sn and $^{132}$Sn nuclei. An
unexpectedly large  contribution of ground state correlations to the
$Q(2^+_1)$ values is found. A non-trivial dependence of the
quadrupole moments of the first $2^+$ states on the neutron excess
is found which can be traced to the negative proton contributions.
A similar behavior could probably be present in other isotope chains.

Quadrupole moments  of unstable nuclei including those near
the exotic $^{100}$Sn and $^{132}$Sn $^{56,78}$Ni are also predicted, which should be of special interest.

\begin{acknowledgement}
The work was partly supported by the DFG and RFBR Grants
Nos.436RUS113/994/0-1 and 09-02-91352NNIO-a, by the Grants
NSh-7235.2010.2  and 2.1.1/4540 of the Russian Ministry for Science
and Education, and by the RFBR grants 11-02-00467-a and 12-02-00955-a.
\end{acknowledgement}

\end{document}